\documentclass[structabstract]{aa}  
\usepackage{graphicx}
\usepackage{natbib}
\usepackage[table]{xcolor}
\usepackage{amssymb}
\usepackage{lscape}
\bibpunct{(}{)}{;}{a}{}{,} 

\begin{document}

   \title{The relation between surface star formation rate density and spiral arms in NGC~5236 (M83)}
 \titlerunning{$\Sigma_{SFR}$ and Arms over NGC~5236}

   \author{E. Silva-Villa
          \and
          S. S. Larsen
          }

   \institute{Astronomy Institute, University of Utrecht,
              Princetonplein 5, 3584 CC, Utrecht, The Netherlands\\
              \email{[e.silvavilla,s.s.larsen]@uu.nl}
         			}
 							

  \abstract 
	{For a long time the consensus has been that star formation rates are higher in the interior of spiral arms in galaxies, 
	compared to inter-arm regions. However, recent studies have found that the star formation inside the arms 
	is not more efficient than elsewhere in the galaxy. Previous studies have based their conclusion mainly on integrated light.
	We use resolved stellar populations to investigate the star formation rates throughout the nearby spiral galaxy NGC~5236.}
	{We aim to investigate how the star formation rate varies in the spiral arms compared to the inter-arm regions, using optical space-based observations of NGC~5236.} 
	{Using ground-based H$\alpha$ images we traced regions of recent star formation, and reconstructed the arms of the galaxy. 
	Using HST/ACS images we estimate star formation histories by means of the synthetic CMD method.}
	{Arms based on H$\alpha$ images showed to follow the regions where stellar crowding is higher. Star formation rates for individual arms over the fields covered were 
	estimated between 10 to 100 Myr, where the stellar photometry is less affected by incompleteness. Comparison between arms and inter-arm
	surface star formation rate densities ($\Sigma_{SFR}$) suggested higher values in the arms ($\sim$0.6 dex). Over a small fraction of one arm we checked how the $\Sigma_{SFR}$
	changes for the trailing and leading part. The leading part of the arm showed to have a higher $\Sigma_{SFR}$ in the age range 10-100 Myr.} 
	{Predictions from the density wave theory of a rapid increase in the star formation at the edge where the stars and the gas enter the density wave are confirmed.
	The $\Sigma_{SFR}$ presents a steep decrease with distance from the center of the arms through the inter-arm regions.}
   \maketitle
%

\section{Introduction}
Studies of OB stars \citep[see e.g.][]{morgan53,mcgruder75,muzzio79,kaltcheva09}, 
show that the concentration of these stars is higher in the Sagittarius arm
of the Galaxy. This suggests an active star formation in the
arms of spiral galaxies. The creation of spiral arms is a problem that has been studied 
since the late 60's \citep[e.g.][]{linshu64,roberts69}. The density wave theory explains how
the spiral arms can be formed and can remain stable over time for an isolated galaxy. The theory predicts
more active star formation in the arms, where the gas compression induced by the density waves triggers the process
\citep[see e.g.][]{linshu64,bashvisser81,knapen92,knapen96,kurtz02,grosbol06}. 
Alternative theories to explain the spiral arms in galaxies have been proposed. 
In the modal theory, inward-moving waves reflect or refract off at the center
of a galaxy (even in the galaxies with a bar in the center), and then the wave comes
back  out as leading or trailing spiral arms \citep[e.g. ][for theoretical and observational 
approaches]{mark74,lau76,bertin89,elmegreen92}. 
Another theory, know as the Stochastic Self-Propagating Star Formation \citep{mullerarnett76}, suggests that 
episodes like supernovae, shock wave or gravitational interactions are responsible to propagate and trigger star formation
\citep[e.g.][for theoretical and observational approaches]{gerola78,seiden79,feitzinger81}. This theory
is very good at explaining flocculent galaxies, while not grand-design ones. It is possible that the whole process
is a combination of these theories. However, in this paper, we will test only results expected from the
density wave theory.

If the density wave theory is correct, the process of gas being compressed by the waves should
lead to many observable effects, e.g. different star formation rates or color-gradients across the arms. 
Assuming a constant angular velocity of the spiral density wave and an approximately flat rotation curve of the stellar component, inside the corotation radius the gas will overtake
the density wave, which will produce an increase in the star formation when compressed. Outside the corotation radius the wave overtakes the gas. 
Consequently, the stars will drift and age, creating a color gradient that can be observed. 
\citet[][]{martinezgarcia09} have studied the color gradient across spiral arms of
13 spiral galaxies. In their work, a reddening free index
was used to study this process, concluding that azimuthal color gradients are common in spiral arms of disk galaxies.
Previous works have tried to relate the spiral arms with the star formation \citep[see e.g. ][]{allen86,tilanus93,stedman01,knapen10,sanchezgil11}.
These studies have shown the relation between the star formation and the spiral structure using different
gas components over different parts of a large set of galaxies (e.g. M83, M51, M100, M101, among others).

Compression of the gas across the galaxy is expected to be observed as regions with enhanced star formation.
The (common) components used to estimate the star formation efficiencies (SFE) in a galaxy are the gas and the
stellar population, the latter being commonly measured through their integrated light. When H$_2$ is use to trace 
star forming regions \citep[e.g.][]{foyle10}, there seemed to be no specific correlation between the SFE 
and the spiral arms. However, studies that estimate the SFE as the fraction of H$\alpha$ to CO and/or HI maps do find a correlation 
between the SFE and the spiral arm \citep[e.g.~][]{lordyoung90,cepabeckman90,knapen92}.
Nevertheless, it is important to note that most of the studies of star formation (either efficiency or rate) and their relation with the arms in spiral galaxies 
have been done using gas components, namely HI, H$_2$, CO, etc combined with SFR tracers (e.g. H$\alpha$),
and unresolved stellar populations, measured through their integrated light \citep[e.g.][]{knapen96,leroy08}.
Tracers based on gas components are assumed to indicate the regions where SFE is high, which could lead to erroneous
estimations, as stated by \citet{foyle10}.

As a new approach, we present in this paper the use of {\em resolved} stellar populations as a tool 
to study how the surface star formation rate density ($\Sigma_{SFR}$) varies in the arms and inter-arm areas of the nearby spiral galaxy NGC~5236.
This galaxy has been selected because it is one of the closest face-on, nearby, grand-design spiral galaxies, where spatial resolution allows a study of this kind. The low
inclination in the line-of-sight reduces the effect of internal reddening. Previous estimations of the $\Sigma_{SFR}$ show values
of $\sim13\times10^{-3}$M$_{\odot}$yr$^{-1}$Kpc$^{-2}$ \citep[e.g.][]{LR00,calzetti_sfr10,silvavilla11}, which is high 
compared to other normal spiral galaxies \citep[e.g.][see table 2]{LR00}. 
A large number of studies have been done in the center of this galaxy, studying different
properties through different wavelengths, all of them suggesting an increasing activity in the star formation
during the last $\sim10$Myr \citep[e.g.][]{ryder95,harris01,houghton08,knapen10}.
Activity in the center of the galaxy combined with the high levels of $\Sigma_{SFR}$ in the disk, show that NGC~5236 is
still actively forming stars, making it ideal to study differences in the star formation and the possible relations with environment and location.

Kinematic studies of NGC~5236 were presented by \citet{lundgren04} based on gas (H$_2$+HI) measurements.
The estimation of the gas surface density in the arms done by Lundgren et al. is higher than the Toomre's 
value for stability ($Q \propto \Sigma^{-1}_{gas}$). This is possibly causing instabilities
in the arms of the galaxy, potentially leading to star formation. However, their estimation of the gas surface density in the inter-arm regions do not show the same high values.
In a further work, \citet{lundgren08} used far-UV, B and H$\alpha$ integrated light to estimate star formation rates, 
while CO was used for gas maps. Lundgren et al. conclude that the star formation presents higher levels
in the nuclear regions, close where the bar ends, and in the arms of the galaxy. The authors also found an increased 
SFE along the arms of this galaxy.

This paper is structured as follows. We introduce the observations, optical and H$\alpha$, used to study the field stellar population and the
regions of recent star formation, respectively, in Sect. 2. Section 3 is devoted to review how the optical bands
were used to run the photometry of the field stars. Using H$\alpha$ (i.e. tracing recent star formation), we will introduce in Sect. 4
the method used to re-construct the arms of the galaxy. Using the photometry from Sect. 3, we
estimated the star formation history of different groups of field stars in Sect. 5. Finally, we present our discussion
and conclusions in Sects. 5 and 6, respectively.

\section{Observations}
We used images from Hubble Space Telescope
and Cerro Tololo observatory, covering different optical wavelengths and H$\alpha$.

\subsection{BVI observations and data reduction}
We use observations of NGC~5236 taken by the {\em Hubble Space Telescope} (HST),
using the {\em Advanced Camera for Surveys} (ACS). The instrument has a resolution of $0\farcs05$ per pixel.
With a distance modulus of  $28.27$ \citep[$\sim4$Mpc,][]{thim03}, 1 pixel corresponds to $\sim$1 pc
in our images.

\begin{figure}[!t]
	\centering
		\includegraphics[width=\columnwidth]{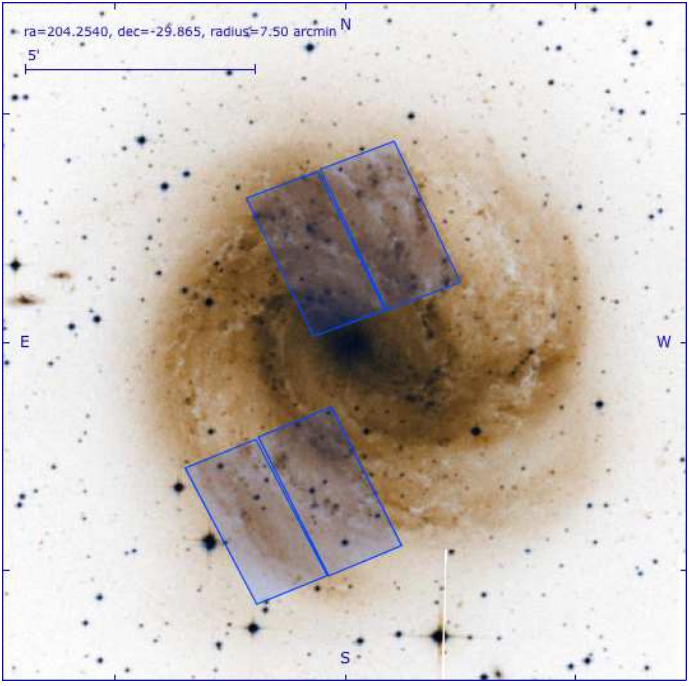}
		\caption{M83 DSS image. Blue lines delineate the ACS pointings used in this paper.}
	\label{fig:m83pointigs}
\end{figure}

The images of this galaxy have been taken in the optical bands F435W ($\sim B$), F555W ($\sim V$), and
F814W ($\sim I$), with exposure times of 680 sec for the bands B and V, and 430 sec for the band I.
Our observations were taken in 2004 as part of Cycle 12, centered at $\alpha:13:37:00$ and $\delta:-29:49:38$ 
and $\alpha:13:37:06$ and $\delta:-29:55:28$ (J2000) for the first and second field observed, respectively.
Figure \ref{fig:m83pointigs} presents a DSS image indicating the regions observed.

The standard STScI pipeline was used for the initial data processing. ACS images were drizzled using the multidrizzle 
task \citep{koekemoer02} in the STSDAS package in IRAF using the default parameters, but disabling the
automatic sky subtraction. Object detection for field stars was performed on an average B, V, and I image, using daofind 
in IRAF.

\begin{figure}[!t]
	\centering
		\includegraphics[scale=0.4]{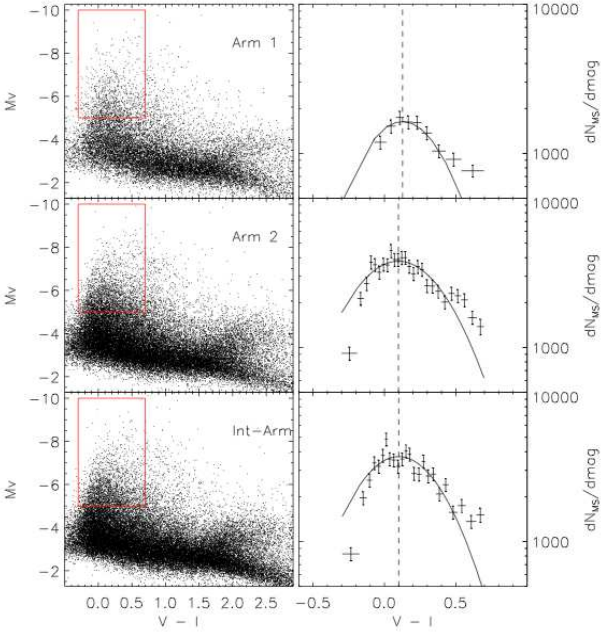}
		\caption{{\em Left column}: Color-magnitude diagrams for the Arm 1, Arm 2, and inter-arm regions. {\em Right column}: Histogram of colors of main sequence stars inside the boxes
		marked over the CMDs. Errors are Poissonian. Vertical dashed lines represent the peak of the Gaussian distributions.}
	\label{fig:cmds}
\end{figure}

\begin{figure}[!t]
		\includegraphics[width=\columnwidth]{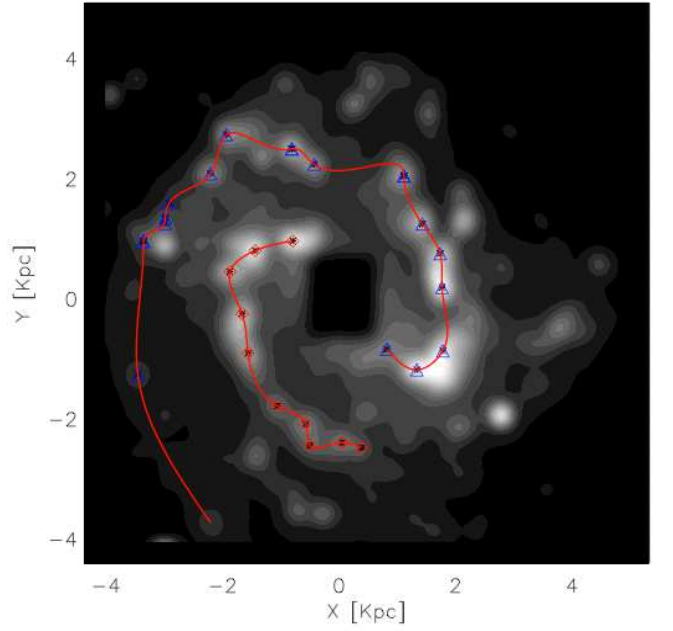}
		\includegraphics[width=\columnwidth]{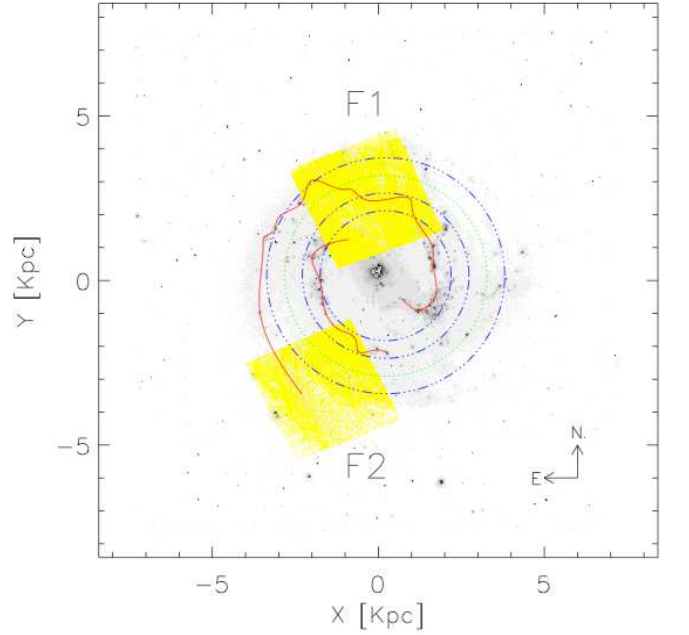}
		\caption{{\em Upper panel}: Blurred image used for the detection of pronounced H$\alpha$ regions. {\em Lower panel}: Original CTIO image of NGC~5236.
			     Overplotted in each panel are the estimated locations of the arms. In the lower panel the two fields observed with 
			     the HST/ACS are marked as yellow regions. Green dot-dashed line is the corotation radius located at 170" \citep{lundgren04}.
			     Dash-dotted lines in blue denote annuli at 200\farcs0, 140\farcs0 and 110\farcs0, see text for details.}
	\label{fig:arms}
\end{figure}

\subsection{H$\alpha$ observations}
We use an archival {\em Cerro Tololo Inter-American Observatory} (CTIO) 1.5m telescope image of NGC~5236, taken in 2006 as part of the {\em Survey for Ionization in Neutral Gas Galaxies} 
\citep[SINGG,][]{singg06}\footnote{SINGG is a subsample of the {\em HI Parkes All Sky Survey} (HIPASS), \citet[][]{hipass04}}.
The survey used the H$\alpha$ and R bands, with a resolution of $0\farcs4$ per pixel.
The total exposure time of the observation for NGC~5236 was 1800 sec, with center coordinates at $\alpha:13:37:02$ and $\delta:-29:52:06$ (J2000).
The image used in this paper is based only on the H$\alpha$ filter. For details on the observations and calibration of the image see \citet{singg06}.

\section{Field stars photometry}
Details of our method of analysis can be found in \citet{silvavilla10}. Below we will reiterate the main points of  the procedures we used to carry out photometry on our data.

Due to the crowding, we performed PSF photometry for field stars.
Using a set of bona-fide stars visually selected in our images, measuring their FWHM with {\it imexam}, we construct
our point-spread function (PSF) using the PSF task in DAOPHOT. This procedure is employed in the same manner for each 
band (i.e. B, V, and I). The PSF stars are selected individually in each band, in order to appear bright
and isolated. PSF photometry is done with DAOPHOT in IRAF. 

Our PSF-fitting magnitudes are corrected to a nominal aperture radius of $0\farcs5$, following 
standard procedures. From this nominal value to infinity, we apply the corrections in \citet{sirianni05}.

HST zero-points\footnote{www.stsci.edu/hst/acs/analysis/zeropoints/\#tablestart} were applied to the PSF magnitudes 
after applying aperture corrections. The zero-points used in this work are 
$ZP_B=25.77$, $ZP_V=25.72$ and $ZP_I=25.52$ magnitudes. Typical errors of our photometry do not change 
dramatically from the ones in \citet[][see its Fig. 2]{silvavilla10}. 

The final color-magnitude diagrams (CMD) for the stars in the arms and the inter-arm regions (see Sect. 5 for definition of the stars that belong to
the arms and in the inter-arm region) are presented in Fig. \ref{fig:cmds}, left column. In the same figure we investigate whether there are any 
indications of differences in the mean extinction from one region to another. We created histograms of the main sequence stars, defined to be stars in the color range
$-0.3\le(V-I)\le0.7$ and in the magnitude range $-10\le M_V\le-5$, as indicated by the red boxes over the CMDs (see Fig. \ref{fig:cmds}, right column). 
The histograms were made with variable bin widths using 100 stars per bin. The number of stars is normalized to the width of the bin. 
We fitted Gaussians to the observed distributions to estimate the maximum of the distributions.
We do not observe a large variation among the Arm 2 and the inter-arm areas, where the peak of the distribution is close the same value ($\sim0.10$$\pm0.02$). However,
the Arm 1 presents a slight shift in the peak of the distribution ($\sim$0.13$\pm0.02$) compared to Arm 2 and inter-arm regions. 
This difference is not large ($\le0.03$, and errors overlap), but we note that
the Arm 1 is closer to the galactic center, where extinction can be affecting the observations. There is also an apparent
increase in the distributions close to $(V-I)\approx$0.5. Inside our photometric errors, separating the main sequence stars from the 
blue He burning phases is not straightforward, but the count of stars can give an indication of the presence of these stars, as seen in Fig. \ref{fig:cmds}, where
at $(V-I)\approx0.5$ the distribution of stars clearly deviates from the Gaussian fit.


\begin{figure}[!t]
	\centering
		\includegraphics[scale=0.4]{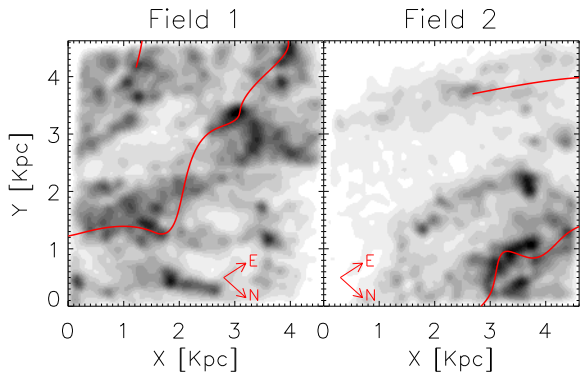}
		\vspace{0.1in}
		\caption{Surface density of stars for the ACS fields. Red lines represent the path of the arms.}
	\label{fig:armsden}
\end{figure}

\begin{figure*}[!t]
	\centering
		\includegraphics[scale=0.55]{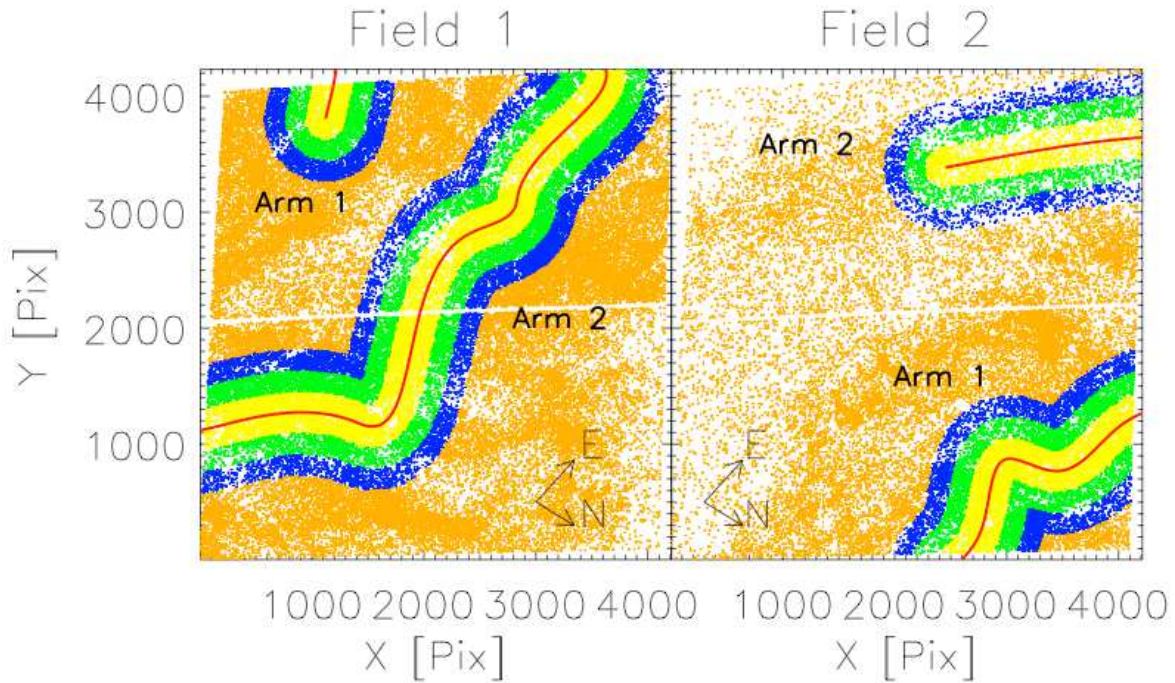}		
		\vspace{0.2in}
		\caption{Selection of the resolved stellar population over the arms of NGC~5236. Red lines show
			     the arms of the galaxy based on H$\alpha$ images. Yellow, green, and blue represent the 
			     selection of stars at different distances that belong to the arms. Orange regions represent the inter-arms.}
	\label{fig:stararms}
\end{figure*}

\section{Defining the spiral arms over NGC~5236}
To identify the arms of NGC~5236, we followed the method
described by \citet{scheepmaker09}, using H$\alpha$ as indicator. The CTIO image
of the galaxy was cropped to remove as much background as possible. Over the new (sub)image,
we use a Gaussian kernel (with a 20 pixels sigma) to blur the image, enhancing the regions
where H$\alpha$ appears to be more concentrated. Because of the high concentration of gas in the 
center of the galaxy, we manually mask this part of the image, which will allow better 
analysis of the regions outside of the center. We analyzed the blurred image with {\em Daofind}
in IRAF to find the places where H$\alpha$ is more concentrated (minimum data value of 400 counts over the background).
The coordinates retrieved by IRAF were visually inspected to remove unnecessary detections, 
i.e. detections that could affect the estimations of the arm's path and that are not part of the
ACS field-of-view. We used a cubic spline interpolation to fit the arms of NGC~5236.
Figure \ref{fig:arms} depicts the CTIO images. The upper panel is the cropped and blurred image,
overplotted with the estimated location of the arms. The lower panel shows the original CTIO image, overplotted 
with the location of the arms (red lines) and the two fields covered by the ACS observations (yellow squares).
Conversion of the coordinates from the CTIO coordinate system to the ACS coordinate system was done using {\em wcsctran} in IRAF. 
First, using the header of the CTIO image, we converted the CTIO coordinate system to WCS coordinates, and then, using
the same procedure, we moved finally to the ACS coordinate system. The different circumferences in Fig. \ref{fig:arms} are marking
the distances 200\farcs0, 140\farcs0, 170\farcs0 and 110\farcs0 used to estimate the velocity of a particle at different radii (see Sect. 6).

We create the density plot of the resolved stellar populations over the ACS fields with the arm's path created using the H$\alpha$.
After overplotting the arms, we observe that the arms follow the region where
the density of stars is higher, as seen in Fig. \ref{fig:armsden}. From the same figure, it is observed
that other regions have high density of stars, e.g. in field 2 there is a possible ``feather'' between
the two arms, and in field 1 we observe many regions with high densities of stars, which is expected, due to
the proximity to the center of the galaxy, where the concentration of gas is higher \citep[e.g.][]{crosthwaite02,lundgren04}.
There is large observational evidence related with ``feathers'' and/or ``spurs'' close (or attached) to spiral arms, which is supported
by theoretical studies \citep[see e.g.][]{scoville01,kim02,shetty06,lavigne06}.

For the rest of this paper we will refer as {\em Arm 1} to the arm that is fully inside the
corotation radius (see green dotted line in Fig. \ref{fig:arms}), while the {\em Arm 2} has one part inside the corotation radius and one part outside of it. Figure
\ref{fig:arms} presents the two arms and the corotation radius, estimated by \citet{lundgren04} to be located at 170".

\section{Selection of stars and star formation histories}
We aim to study here how the star formation varies from the center of the spiral arms to the inter-arm regions.
We use the resolved stellar population
detected over our ACS fields and estimate their star formation histories at a fixed distance from the center of the arm.

\subsection{The arms}
Stars were selected every 0.2 Kpc, assuming that the arms presented above mark the "center" of the distribution.
Figure \ref{fig:stararms} shows the distribution of the stars which follow the arms at the distances 0-0.2, 0.2-0.4, and 
0.4-0.6 Kpc, represented with the colors yellow, green, and blue, respectively. In our second field
the distance between the two arms would allow to cover a larger area, however, we kept the same distances as
in the first field in order to have comparable results among the fields.

For each of the selected group of stars, star formation histories (SFH) were calculated using the synthetic CMD method \citep{tosi91}.
A description and tests of the IDL-based program used to estimate the SFH is given in \citet{silvavilla10}. For more applications, see \citet{silvavilla11}.
We have updated the code presented by \citet{silvavilla10} allowing the use of multiple metallicities. When using multiple metallicities, the code models the Hess
diagram as a weighted sum of all isochrones in the input library. The final SFH is obtained adding together the individual SFR over time
for each assumed metallicity.

\begin{figure}[!t]
	\centering
		\includegraphics[scale=0.4]{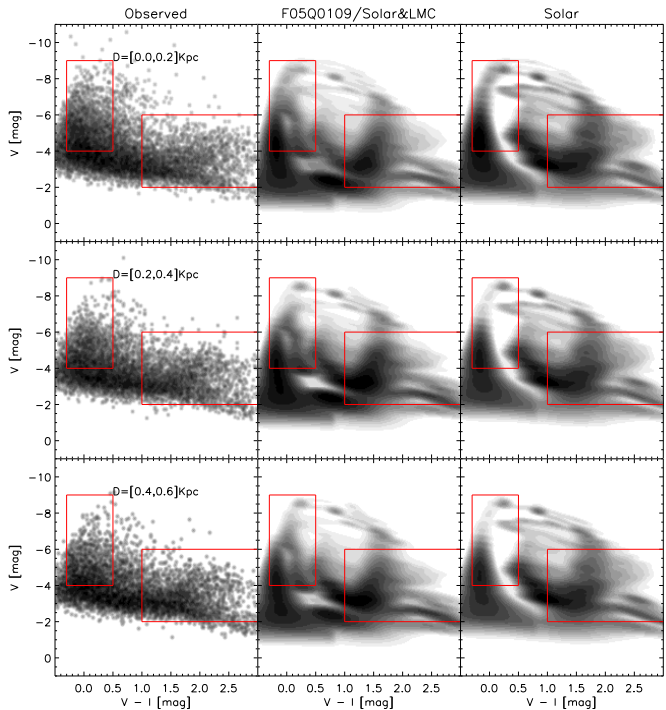}
		\caption{{\em Left colum}: CMDs for the Arm 2 in the field 1 at different distances from the center of the arm.
			      {\em Middle colum}: Best fit CMDs assuming solar and LMC like
			      metallicities, a binary fraction of 0.5 with a mass ratio between 0.1-0.9, and a fixed A$_B$ foreground extinction of 0.29. {\em Right colum}: Best fit CMDs assuming solar 
			      metallicity, a binary fraction of 0.5 with a mass ratio between 0.1-0.9, and a fixed A$_B$ foreground extinction of 0.29.
			      The fits were done using the red boxes shown in the figure.}
	\label{fig:hessfits}
\end{figure}

The parameters used for the estimation of the SFH in NGC~5236 are: Distance modulus of 28.27, solar and LMC-like metallicities, a binary fraction of
0.5 with a mass ratio between 0.1-0.9 (assuming a flat distribution and no binary evolution), and using the color combination V-I.
We normalized our estimations by the areas covered, having then the surface star formation rate density ($\Sigma_{SFR}$
[M$_{\odot}$yr$^{-1}$Kpc$^{-2}$]). The estimation of the different areas covered was carried out following a similar procedure as for the
selection of the stars. We create an image of the size of the ACS fields and calculate the distance of each pixel to the arms. Having the total amount of
pixels under the desired area, the areas were calculated following the relation:
\begin{equation}
A_i=A_{total} \times \frac{Npix_{A_i}}{Npix_{A_{total}}} \ ,
\end{equation}
where A$_i$ is the area to be calculated, A$_{total}$ is the total area covered by the ACS (21.3 Kpc$^2$, each field), and Npix$_{A_i}$ and Npix$_{A_{total}}$ are
the number of pixels for the area {\em i} and the total number of pixels, respectively. 

Figure \ref{fig:hessfits} illustrates the fits done for field 1 for the three areas covering Arm 2. We note that the fits are far from perfect. 
Trying to understand what could be the cause of the mismatch between observed and modelled CMDs, we repeated the fits 
varying the input parameters. As an example, the right column in Fig. \ref{fig:hessfits} 
shows the fits using only solar metallicity (instead of solar and LMC). It is clear that solar metallicity leaves a "gap" between the main sequence
stars and the blue loop, which is not seen in our observed CMDs (first column). 
The combination of both metallicities gives a better fit to the data, as the blue loops extend to higher effective temperatures (bluer colours) for LMC-like metallicity, and we
can not reject the possibility that higher metallicities are present. Regardless of the improvement when using a combination of metallicities, 
we note that neither red nor blue He burning phases are well fitted by our program,
appearing bluer and redder, respectively, in our final fits, when compared with the observations.
Assumptions of no binaries did not show any improvement. Following
the small change in extinction, as suggested in Sect. 2, we used a time dependent extinction, but
we did not see any improvement. 
The middle column presents the best fit which combines the parameters described in the previous paragraph.

In view of the difficulties reproducing the observed CMDs, caution should clearly be exercised when interpreting SFHs inferred from these fits.

\begin{figure*}[!t]
	\centering
		\includegraphics[scale=0.5]{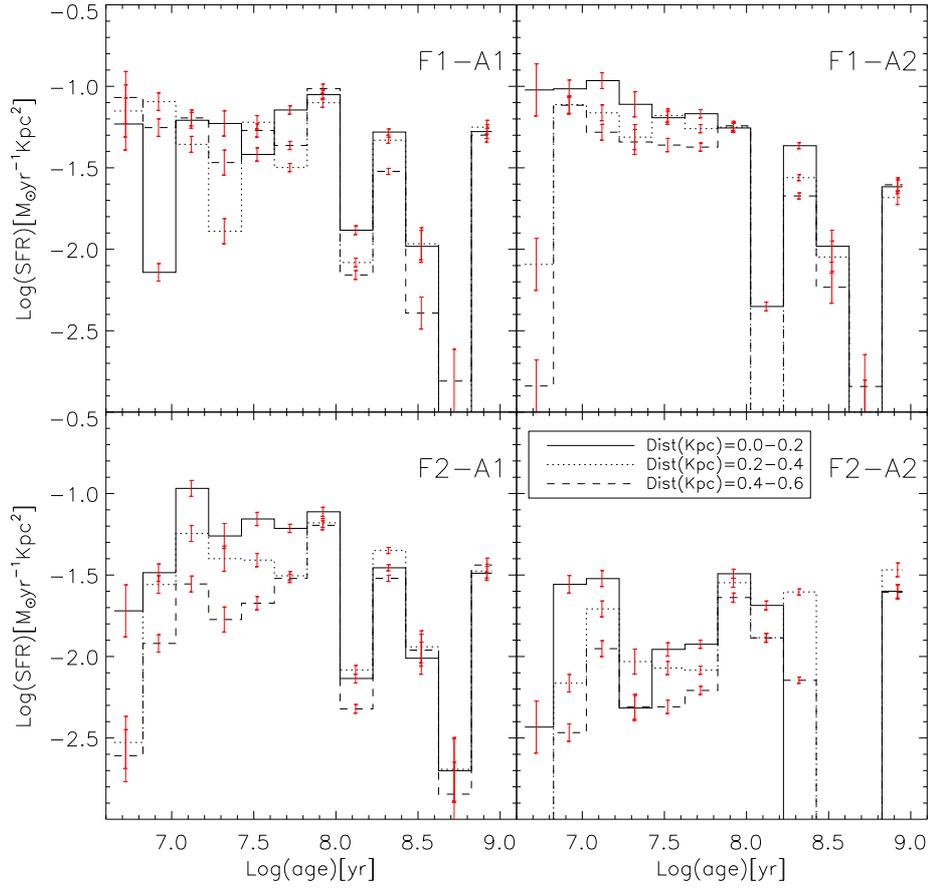}
		\caption{Star formation histories for each arm over each field. F\#-A\# refers to the field and the arm estimation. Errors are the random errors \citep[see][]{silvavilla10}.}
	\label{fig:sfhfields}
\end{figure*}
\begin{figure*}[!t]
	\centering
		\includegraphics[scale=0.4]{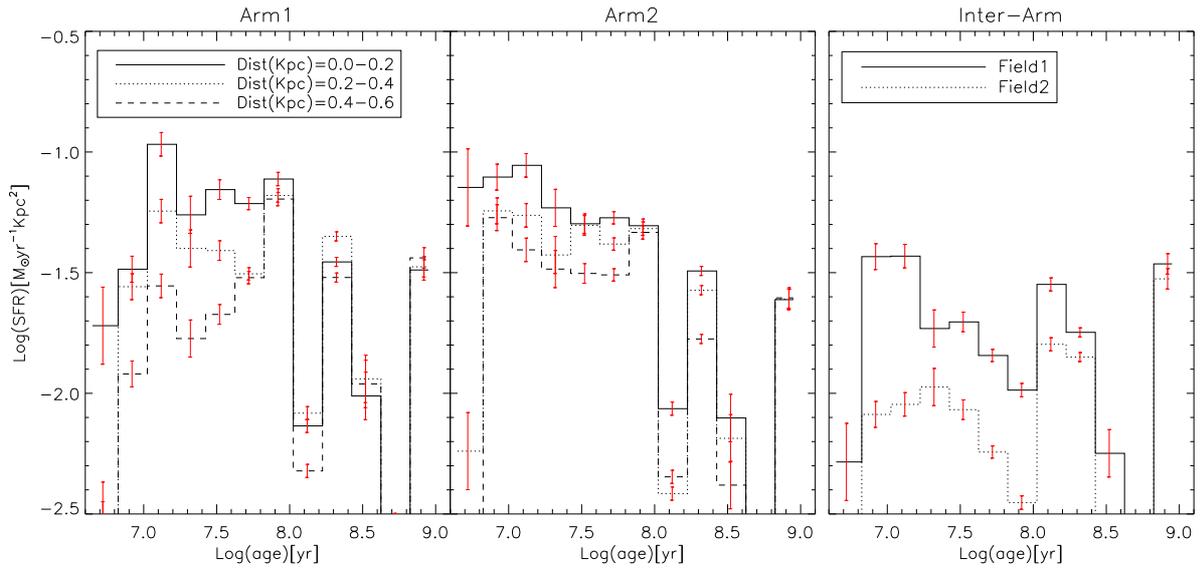}
		\caption{{\em Left/Middle panels}: Star formation histories for each arm after joining fields. 
			      {\em Right panel}: Star formation history of the inter-arm areas. Errors are the random errors \citep[see][]{silvavilla10}.}
	\label{fig:sfharms}
\end{figure*}

For each field and each arm, between 10 and 100 Myr (age range less affected by incompleteness), 
we estimated the mean values for $\Sigma_{SFR}$. We also combined results for each arm, adding the results from each field,
see Fig. \ref{fig:sfhfields} and \ref{fig:sfharms}. We summarize our results in table \ref{tab:sfhs}.

Independent measurements of $\Sigma_{SFR}$ (i.e. per field, per arm, per area covered) are shown in Fig. \ref{fig:sfhfields}, 
where Arm 2 suggest that during the past 100 Myr there has been an increase in the $\Sigma_{SFR}$ (right panels). 
As the individual SFHs are quite noisy, we joined the results for all areas of Arm 2, but for
the Arm 1 we only used the results from the field 2, because the areas covered in the field 1 were too small 
and a clear separation between the arm and the inter-arm region was not easy to performed, compromising the results. The results are presented
in Fig. \ref{fig:sfharms}. From this figure it becomes clear that there is an increase in the $\Sigma_{SFR}$ between 10 to 100 Myr for 
Arm 2. 
This increase is observed for each one of the areas covered in that particular arm, which suggest a physical effect. 
In Sect. 6 we interpret this result.

\subsection{The inter-arm regions}
To estimate the $\Sigma_{SFR}$ for the inter-arm regions we used the stars outside the arms (orange regions in Fig. \ref{fig:stararms}).
Stars selected in this way were passed to our program and SFHs for each field
were estimated (see Fig. \ref{fig:sfharms} right panel). The areas were calculated as the total area (A$_{total}$) minus the areas covered by the arms, then
having an estimation of the $\Sigma_{SFR}$ (see table \ref{tab:sfhs}). We added the independent estimations for the two fields and
plotted the mean value in Fig. \ref{fig:sfhdist} as a continuum value, which is lower than the values derived for the arms,
making a clear distinction between the SFH's in the arms compared to the inter-arm areas.
It is important to note the differences among the fields. The first field, which is close to the center
of the galaxy displays a higher $\Sigma_{SFR}$, even suggesting a small increase over time. The second
field, which is located further away from the center, displays a lower $\Sigma_{SFR}$. This separation
indicates differences that could be related with the amount of material that could potentially be used
to form stars. It is expected to find more gas concentrated in the center of the galaxy than in the
outer parts \citep[e.g.][]{crosthwaite02,lundgren04}. 
Also important to note is the increase between $\sim$20-100 Myr. This
could be related to the limits adopted for the arms and inter-arm regions, where the increase in 
the number of stars at the "edge" of the arms could be contaminating the inter-arm areas.
Also, as noted in the previous sections, there are regions in the inter-arms that present high levels of H$\alpha$,
suggesting recent star formation activity. Nevertheless, keeping in mind the discussion in the previous section, the 
increase could be an artifact of the fitting process.

\begin{table}[!t]
\centering
\caption{Surface star formation rate densities at different distances from the center of the arms. Errors are the standard deviation of the mean. Arm 1 values are for field 2 only.}

\begin{tabular}{c c c c }
\hline \hline
Field\_Arm & Dist$_{arm}$ & Area & $\Sigma_{SFR}$ \\ 
 & [Kpc] & [Kpc$^2$] & [$\times 10^{-3}$M$_{\odot}$yr$^{-1}$Kpc$^{-2}$] \\ \hline
\multicolumn{4}{c}{Arms covered} \\
Arm1 &  0.0-0.2  & 1.06 & 67$\pm$24 \\
Arm1 &  0.2-0.4  & 1.05 & 43$\pm$14 \\
Arm1 &  0.4-0.6  & 1.01 & 29$\pm$18 \\
Arm2 &  0.0-0.2  & 3.29 & 67$\pm$15 \\
Arm2 &  0.2-0.4  & 3.44 & 41$\pm$19 \\
Arm2 &  0.4-0.6  & 3.54 & 31$\pm$17 \\ \hline

\multicolumn{4}{c}{Inter-arms  covered} \\
Field1 &  ---  & 12.84 & 23$\pm$11 \\
Field2 &  ---  & 15.24 & 8$\pm$3 \\ \hline

\label{tab:sfhs}
\end{tabular}

\end{table}

\section{Discussion}
The $\Sigma_{SFR}$ of the arms in NGC~5236 estimated here suggest that the 
arms of the galaxy have increased their star formation rate during the last 100 Myr by $\approx$0.2 dex (see Fig. \ref{fig:sfharms}, left and middle panels). 
In the right panel of the same figure we present the $\Sigma_{SFR}$ of the inter-arms, which presents three important features: 
(1.) There is a clear difference in the star formation history among the two fields, where
the field 1 presents a higher $\Sigma_{SFR}$ in comparison with the field 2 (see table \ref{tab:sfhs});
(2.) there is an apparent increase in the star formation history, which could be due
to different factors, e.g. feather/spurs with active star formation or small differences in
the separation between arm and inter-arm areas; and
(3.) both fields present a mean $\Sigma_{SFR}$ lower than in the arms areas (see table \ref{tab:sfhs}).

\begin{figure}[!t]
	\centering
		\includegraphics[height=3in,width=\columnwidth]{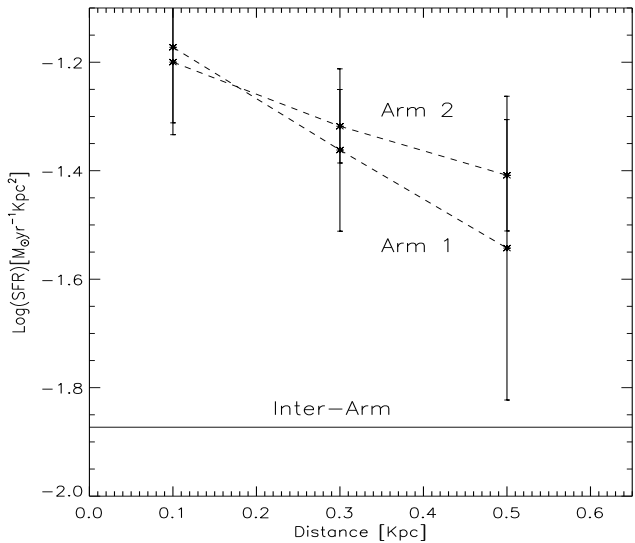}
		\caption{$\Sigma_{SFR}$ as a function of distance for both arms. Straight horizontal line represents the mean $\Sigma_{SFR}$ for the inter-arm region.}
	\label{fig:sfhdist}
\end{figure}

Figure \ref{fig:sfhdist} shows the variation of the mean $\Sigma_{SFR}$ (between 10 and 100 Myr) as a function of 
distance from the center of the arms, and the $\Sigma_{SFR}$ for the inter-arm regions, as defined in the previous section.
Starting from the innermost region of the arms, traced by H$\alpha$,
we observed a very fast and steep decrease moving outward through the arms. The estimations
of $\Sigma_{SFR}$ show a clear difference between the arms and the inter-arm regions, the former having 
a higher value in the center (by $\sim0.6$ dex), while reaching similar levels at the external regions, where the separation between arm and inter-arm becomes dificult.
In Fig. \ref{fig:armsden} we showed that H$\alpha$ traces the regions
where the crowding of stars is higher, suggesting that the recent star formation is higher in the arms of
the galaxy. There are other regions in the same figure where large concentrations of stars
are present, but based on the procedures described here, in those regions H$\alpha$ could be interpreted as ``feathers'' and/or ``spurs'' based on our procedures. We do not disregard
the possibility of having recent star formation in the inter-arm regions, possibly induced by e.g. supernovae explosion.
\citet{lundgren08} suggested that in the bar and in the arms of
M83, the $\Sigma_{SFR}$ and the SFEs are higher than in the inter-arm areas.
Our measurements of $\Sigma_{SFR}$ showed to be higher in the arms of the galaxy, in agreement with the work done by \citet{lundgren08}, when
compared to the inter-arm regions (at least qualitatively), however, we do not draw any conclusion regarding the bar of the galaxy. 

We tested results expected from the density wave theory studying the $\Sigma_{SFR}$ in the fields covered in this paper.
Using increasing distances from the center of the arms (as defined in this paper through H$\alpha$)
a comparison was made to see the spatial variations in the arms of the galaxy, looking for the differences
between the arms and the inter-arm regions.

\begin{figure}[!t]
	\centering
		\includegraphics[width=\columnwidth]{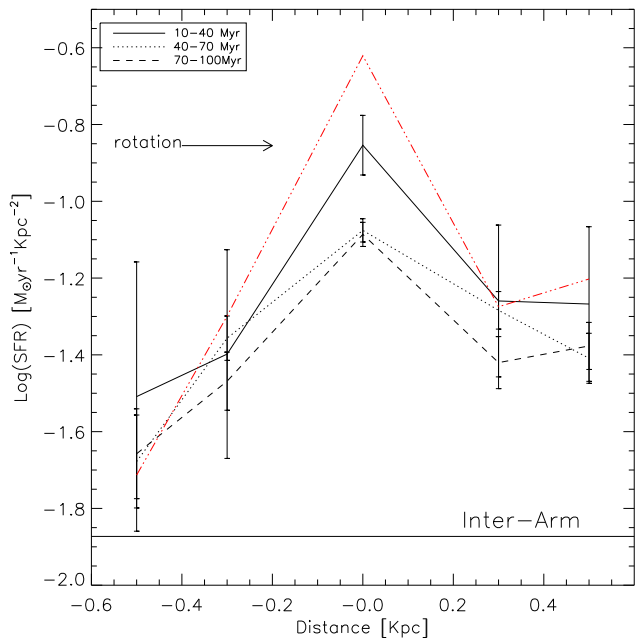}
		\caption{Spatial variation of $\Sigma_{SFR}$ across the Arm 2, observed in field 1. The values are averaged over age every 30 Myr. Errors are the standard deviation of the mean.
			      Red dash-dotted line represent the number of massive, bright stars in the main sequence normalized by the are, see text for details.}
	\label{fig:sfhacross005}
\end{figure}

\begin{figure}[!t]
	\centering
		\includegraphics[width=\columnwidth]{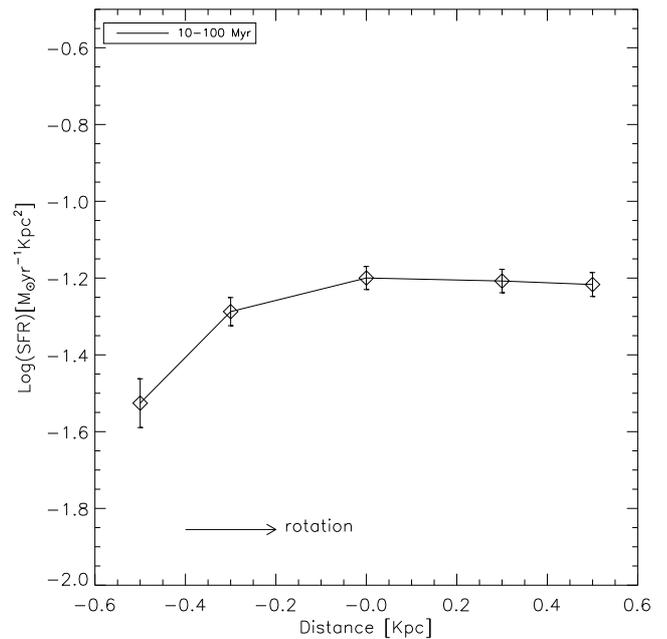}
		\caption{Variation of $\Sigma_{SFR}$ across the Arm 2, observed in field 1. Variation averaged over the whole age range 
			     10-100 Myr. Errors are the standard deviation of the mean.}
	\label{fig:sfhacross}
\end{figure}

The density wave theory \citep{linshu64,roberts69} explains how
it is possible to have semi-stable arms in spiral galaxies. We test predictions from this theory using resolved stellar population.
If the gas is overtaken by the density wave (or viceversa, depending on the radius), then,
it is expected to observe higher star formation rate per unit area at the edge where the gas enters the density wave 
due to the compression of the gas \citep[e.g.][]{allen85,elmegreen11}. 
Also, from the density wave theory it is expected to observe young stars
close to the regions where the gas is more concentrated (i.e. where the star formation is
happening), but with time and rotation, stars evolve and travel through the arm.
This process can be observed as a color-gradient across the arms. Theoretical predictions
of the time ranges for color-gradients were estimated by \citet{martinezgarcia09} using the reddening-free
parameter $Q(rJgi)$ (where $r, J, g, i$ are the photometric bands used). Mart\'{i}nez-Garc\'{i}a et al. show that the maximum of $Q$ is reached
at $\sim2\times10^7$yr after a burst of stars (assuming a \citet{salpeter55} IMF), and then starts to decline,õ reaching lower values at $\sim100$Myr after the burst. This estimate is
of course affected by inclination, the photometric bands used, metallicity, etc.
\citet{martinezgarcia09} studied the color gradient variation over different spiral galaxies. The authors used
a sample of 13 spiral galaxies, and showed how a fraction of the arms analyzed clearly showed the expected color gradient suggested by the density wave theory.
We used resolved stellar populations over the Arm 2, in field 1, where the statistics are better (see Fig. \ref{fig:stararms}), to observe
the variations of the $\Sigma_{SFR}$ inside this (part of the) arm.
Separating the areas at 0.2-0.4 Kpc and 0.4-0.6 Kpc into "trailing" and "leading" parts of 
the arm (left and right in Fig. \ref{fig:stararms}), we estimated $\Sigma_{SFR}$
separately.  Using the rotation curve found in \citet{lundgren04}, we estimated how much time
a particle will need to move 0.2 Kpc at different radii. Inside corotation radius \citep[R$_{CR}=170"$,][$\sim$3 Kpc for the distance modulus assumed in this paper]{lundgren04},
the time for a particle to move that distance is close to $\sim$5 Myr. Outside corotation, the time is increased
by a factor of $\sim$3 times. As observed in Fig. \ref{fig:arms}, the circumference at different radii are not
crossing perpendicular to the arms. Because of this, the real path a particle will cross is larger than 0.2 Kpc. We
decided to estimate mean $\Sigma_{SFR}$ every 10Myr, and check  how it varies with time. Figure \ref{fig:sfhacross005} 
shows the $\Sigma_{SFR}$ as a function of position in the arm for different time intervals.
Inside corotation, the gas overtakes the density wave, creating a very steep increase in the first  few Myr, which must start to decrease with time, as observed
in the figure. Fig. \ref{fig:sfhacross} shows the average of the $\Sigma_{SFR}$ over an age range between 10-100Myr.
As indicated by the arrow, and being (mostly) inside corotation, a very steep increase in the $\Sigma_{SFR}$ 
is observed (by a factor of $\sim$0.4 dex), reaching a maximum where the location of the H$\alpha$ is denoting the center of the arm, and then showing
a decrease with time. The small fraction of the arm outside corotation was not analyzed separately due to the low statistics
and the lack of smoothness of the arm at that location (see Fig. \ref{fig:arms}).

As a consistency check to our results, we selected the massive bright stars over the CMD for Arm 2 in field 1, defined as stars
between $-0.3\le (V-I) \le 0.7$ and $-10\le M_V \le -7$. The number of stars in each area were divided by the area covered and then passed to logarithmic scales.
The estimated stellar densities were shifted to the same levels of the SFH estimations for the comparison (see Fig. \ref{fig:sfhacross005} red dash-dotted line).
We observed that the distribution of these stars is in excellent agreement with the SFRs derived from the fits to the CMDs. 

These results are in good agreement
with the conclusions of \citet{martinezgarcia09} for spiral galaxies, and with the predictions of the density wave theory, which
states that the leading part of the arm should have a higher star formation compare to the trailing part of an arm, inside the corotation radius.

\section{ Summary and conclusions}
Using H$\alpha$ CTIO images of the galaxy NGC~5236, we have traced the arms of the galaxy. Superb resolution of HST/ACS observations have
allowed us to identify the stellar populations, which we have used to observe the variations of $\Sigma_{SFR}$ in the inner
parts of the arms and compare how this parameter behaves with increasing distance from the center of the arms through
the inter-arm regions. 

H$\alpha$ showed to be a good tracer of higher concentrations of stars across the regions covered with our HST observations.
Star formation histories were estimated using the synthetic CMD method \citep{tosi91} across the arms and the
inter-arm regions separately. However, the fits to the CMDs done are far from perfect. Verifying different
parameters did not further improved our results, however we observed that the inclusion of more than one
metallicity helps to reproduce better the observed CMDs.
We observed that the arms of NGC~5236 present a higher $\Sigma_{SFR}$, compared to the mean value for the inter-arm regions combined by $\approx$0.6 dex.
Our analysis of resolved stellar populations leads to similar conclusions as \citet{lundgren08} for this galaxy, which were based on gas,
dust, and integrated light studies.

Comparison of the trailing and leading part of the Arm 2 (only for the observations in field 1) indicates that the region leading has a higher
$\Sigma_{SFR}$. This result is in good agreement with \citet{martinezgarcia09}, who used a reddening-free photometric
index ({\em Q}) to trace the variation of the stars formation across a spiral arm, using a sample of 13 spiral galaxies.
Both results are in agreement with the density wave theory, which suggested that the region hitting, either the gas
or the spiral density wave, must induce a higher star formation due to the increase in the density of the gas.

An alternative to the spiral density wave theory is the theory of stochastic self-propagating star formation, which is quite successful at explaining flocculent galaxies \citep{mullerarnett76}. In this theory, star formation propagates through shock waves produced by supernova events. These shock waves induce new regions of active star formation, which combined with the differential rotation of the galaxy, create the fragmentary spiral patterns observed in flocculent galaxies. However, it is unlikely that this scenario can account for grand-design spiral galaxies such as M83. It is possible that supernova shocks and other feedback leads to some propagation of star formation within the spiral arms, but our observations are not well suited for addressing this issue as we cannot easily age date individual stars at the very young ages involved. Also, our broad-band imaging does not allow us to observe the detailed structure of the gas, which could be a good indicator of this process.


\begin{acknowledgements}
We would like to thank the referee, Diederik Kruijssen and Juan Carlos Mu{\~n}oz-Cuartas for insightful comments and discussions that helped improve this article.
This work was supported by an NWO VIDI grant to SL.
\end{acknowledgements}

\bibliographystyle{aa}
\bibliography{../../../article_set}

\end{document}